\documentclass[twocolumn,secnumarabic,amssymb,nobibnotes,aps,longbibliography,pra,showpacs,superscriptaddress]{revtex4-1}

\usepackage{bm}
\usepackage{amsfonts,epstopdf,times}
\usepackage{epsfig}
\usepackage{amsfonts,subfigure}
\usepackage{dcolumn}
\usepackage{bm}
\usepackage{graphicx,times,psfrag,color,hyperref,xcolor}
\usepackage{amssymb,latexsym,mathrsfs,amsmath,amsthm}
\usepackage[utf8]{inputenc}
\usepackage[mathlines]{lineno}
\usepackage{verbatim}
\usepackage{lipsum}
\usepackage[percent]{overpic}
\usepackage{orcidlink}
\usepackage{algorithm}
\usepackage{algpseudocode}
\usepackage{cuted, mathtools}
\usepackage{tikz}
\usepackage{multirow}

\usepackage{braket}
\usepackage{bbm}
\DeclareMathOperator{\Tr}{Tr}

\newcommand{\id}{\mathbbm{1}}

\newtheorem{definition}{Definition}
\newtheorem{theorem}{\bf{Theorem}}

\newtheorem{lemma}{\bf{Lemma}}

\begin{document}

\title{Circuit Design for a Star-shaped Spin-Qubit Processor via Algebraic Decomposition and Optimal Control}

\author{Yaqing X. Wang \orcidlink{0000-0003-1457-960X}}
\affiliation{Forschungszentrum Jülich, Institute of Quantum Control,
	Peter Grünberg Institut (PGI-8), 52425 Jülich, Germany}
\affiliation{Institute for Theoretical Physics, University of Cologne, 50937 Köln, Germany}

\author{Tommaso Calarco \orcidlink{0000-0001-5364-7316}}
\affiliation{Forschungszentrum Jülich, Institute of Quantum Control,
	Peter Grünberg Institut (PGI-8), 52425 Jülich, Germany}
\affiliation{Institute for Theoretical Physics, University of Cologne, 50937 Köln, Germany}
\affiliation{Dipartimento di Fisica e Astronomia, Universit\`{a} di Bologna, 40127 Bologna, Italy}

\author{Felix Motzoi \orcidlink{0000-0003-4756-5976}}
\affiliation{Forschungszentrum Jülich, Institute of Quantum Control,
	Peter Grünberg Institut (PGI-8), 52425 Jülich, Germany}
	\affiliation{Institute for Theoretical Physics, University of Cologne, 50937 Köln, Germany}

\author{Matthias M. Müller \orcidlink{0000-0002-1428-9680}}
\email{ma.mueller@fz-juelich.de}
\affiliation{Forschungszentrum Jülich, Institute of Quantum Control,
	Peter Grünberg Institut (PGI-8), 52425 Jülich, Germany}

\begin{abstract}
As quantum processing units grow in size and precision we enter the stage where quantum algorithms can be tested on actual quantum devices. To implement a given quantum circuit on a given quantum device, one has to express the circuit in terms of the gates that can be efficiently realized on the device.
We propose an algorithm based on algebraic circuit decomposition for tailored application of optimal-control gates for quantum computing platforms with star-shaped topologies. We then show numerically how the resulting circuits can be implemented on a quantum processing unit consisting of a nitrogen-vacancy center in diamond and surrounding nuclear spins.
\end{abstract}

\maketitle

\section{Introduction}
Quantum algorithms at the noisy intermediate scale
~\cite{RevModPhys.94.015004, Preskill2018} such as approximate \cite{farhi2014quantumapproximateoptimizationalgorithm, farhi2019quantumsupremacyquantumapproximate, kazi2024analyzingquantumapproximateoptimization} or variational algorithms~\cite{BravoPrieto2023variationalquantum, peruzzo_variational_2014}, or the quantum Fourier transformation~\cite{aqft,cleve2000fast,hales2000improved,Chen2020}, quantum simulation~\cite{Lloyd1996, Aspuru-Guzik2005, Georgescu2014, Ruh2023} or algorithms for quantum error correction \cite{Finsterhoelzl2022,Terhal2015}  have gained much interest in recent years. The compilation of these algorithm as circuits on the available quantum devices is an important step toward implementation and an active field of research with compilation methods based on, e.g., numerical optimization or machine learning~\cite{qiskit2019, Younis2021, Jones2022, Kreppel2023, Preti2024}. For universal quantum computing, one needs a certain minimal gate set, where different choices are possible and can depend on the physical platform~\cite{NC}. The universality of a gate set can be shown, e.g., through algebraic decomposition~\cite{Khaneja2000, Zhang2003b}, which in turn can be used also to decompose a circuit into a given gate set, i.e., for circuit compilation~\cite{Zhang2004a, Nakajima2003, Drury2008}. Once the circuit is compiled, the single gates have to also be implemented with high fidelity. This can be achieved, e.g., by methods of quantum optimal control (QOC)~\cite{Glaser2015, brif2010control, Rembold2020, crab,Koch2022} or dynamical decoupling~\cite{Viola1999,Casanova2015}. Numerically optimized pulses can then be stored in a pulse library, which can be made more versatile toward a continous gate set by combining QOC with tools of machine learning~\cite{DeKeijzer2024,Preti2022}. One can also combine QOC with algebraic decomposition to exploit symmetries in the system and to be able to focus on the relevant degrees of freedom in the system, such as a given local equivalence class of gates or the property of being a perfect entangler~\cite{Mueller2011, Watts2015, Goerz2015, Yuan2015, Goerz2017}. The resulting gates can then be the starting point of the aforementioned circuit compilation so that the entire circuit requires for instance only the perfect entangler obtained by QOC together with local gates.

In this work, we study in particular quantum circuits for central spin models~\cite{zhang2020efficient,Chen2020, Finsterhoelzl2022, Ruh2023} like the nitrogen vacancy (NV) center in diamond with its surrounding nuclear spins.
NV centers in diamond~\cite{jelezko2004observation,Hanson2006,Wrachtrup2006, waldermann2007colorcenter,maze2011properties,neumann,van2012decoherence,Schirhagl2014,gali2019ab,Rembold2020}~are a promising candidate for room-temperature quantum processing units. Together with their surrounding nuclear spins, they form a spin register with long coherence time~\cite{Cappellaro2009, Neumann2010, weber2010quantum, neumann, Chen2020, Unden2016, liu2018quantum, Joas2024} where the nuclear spins can also serve as quantum memory~\cite{Fuchs2011} and the NV center allows for optical (even single-shot) read-out~\cite{Neumann2010a, Joliffe2024}. Universal quantum gates have been shown in this system at high fidelity~\cite{rong2015experimental,Joas2024,Vetter2024,Bartling2024}, allowing for the demonstration of quantum error correction protocols~\cite{Waldherr2014,Unden2016}. These universal high-fidelity gates can be implemented via dynamical decoupling \cite{Casanova2015,Casanova2016, Casanova2017, Tratzmiller2021} or QOC~\cite{Rembold2020,Chen2020,Vetter2024}.

 In particular, here, we present an algorithm for circuit decomposition based on the Cartan decomposition of $SU(2^N)$. Such an algorithm was already proposed in Ref.~\cite{Nakajima2003} for partial decomposition, and Ref.~\cite{Drury2008} proposed a further step toward recursive decomposition but did not provide a corresponding computing algorithm. This work provides the algorithm and connects explicitly such decomposition to the innate Lie algebra structure of centrally-coupled qubit systems resulting from the interaction of the central spin with each of the register spins. We show how to implement specific two-qubit gates between the nuclear spins activated via the central NV center electronic spin, as well as the Quantum Fourier Transformation (QFT)~\cite{cleve2000fast,hales2000improved,Chen2020}. We then numerically optimize the pulses that generate the necessary gates and simulate the time-dynamics of the circuits.
 
 The paper is organized as follows. In Section~\ref{section:NV_section} we introduce the model of our quantum processing unit and the basic gate schemes. In Section~\ref{section:cartan} we introduce the algorithm for algebraic decomposition of a quantum circuit into the basic gates. In Section~\ref{section:gate_design} we show how one can generate the control pulses needed to implement the individual gates of the circuits, and in Section~\ref{sec:numerical_implementation} we describe the method and relevant parameters of our numerical simulations. The results are presented in Section~\ref{section:results} and discussed in Section~\ref{sec:discussion}. We finally conclude and give an outlook in Section~\ref{sec:conclusions}.

\section{System}
\label{section:NV_section}
Following Ref.~\cite{Waldherr2014,Chen2020}, we consider an NV-center based quantum processor with the NV center as a central electron spin-1 system coupled to $N$ spin-1/2 nuclear spins that serve as qubit register, in particular the $^{15}$N nuclear spin and  $N-1$ surrounding $^{13}$C nuclear spins.
Two magnetic fields are present in the considered set-up. Firstly, a strong background static magnetic field $B_0$ is aligned with the NV center axis.
Secondly, a control magnetic field $B_1(t)$ aligned along the $\hat{x}$ axis of the system allows to drive the excitations in the system. The frequencies that trigger electronic transitions on the NV center are in the microwave domain, vastly detuned from the range of transition frequencies on nuclear spins, which are in the radio frequency domain. This domain separation makes individual addressing of the electron spin and the nuclear spins possible as we will see in more detail below.

\noindent Prior to further simplifications, the Hamiltonian of the system can be written as~\cite{Rembold2020}:

\small\begin{gather}
	\hat{H}(t) = \overbrace{
		\hbar D \left[ S_z^2 - \frac{2}{3} \right]  
	}^{\text{zero-field splitting}} + \overbrace{ 
		\hbar\gamma_{NV}( B_1(t) S_x  + B_0 S_z)
	}^{\text{electronic Zeeman}} \nonumber\\
	+ \hbar \sum_{i=1}^N \left( \underbrace{
		\overrightarrow{S}\overleftrightarrow{A}_i \overrightarrow{I}_i
	}_{\text{hyperfine}} - \underbrace{
		\gamma_i  (B_1(t) I_{i,x}  + B_0 I_{i,z})
	}_{\text{nuclear Zeeman}} + \underbrace{
		Q_i I_{i,z}^2
	}_{\text{nuclear quadrupole}}
	\right).
\end{gather}
Here, $D$ is the zero-field splitting, $\gamma_{NV}$ is the gyromagnetic ratio for the NV electron spin, $\gamma_i\in \{\gamma_{~^{15}N},\gamma_{~^{13}C}\}$ is the gyromagnetic ratio for the nuclear spins, $\overrightarrow{S}=({S}_x, {S}_y, {S}_z)$ is the electron spin-1 operator, $\overrightarrow{I}_i=(I_{i,x},I_{i,y},I_{i,z})$ are the spin-1/2 operators of nuclear spin $i$, $\overleftrightarrow{A}_i$ is the hyperfine-interaction tensor for nuclear spin $i$, and $Q_i $ is the quadrupole moment of nuclear spin $i$. From now on we set $\hbar=1$.
We assume that the energy splitting by $B_0$ together with $D$ is large enough such that the secular approximation can be made~\cite{Schirhagl2014, maze2011properties,Chen2020}. As a consequence, only the $z$-component of the spin state of the NV center is to be considered in the spin interaction with the nuclei, such that $\overleftrightarrow{A}_i$ can be described by its $A_{i,zx}$, $A_{i,zy}$ and $A_{i,zz}$ entries. We furthermore assume that the perpendicular couplings are small, allowing for an approximate $ZZ$ coupling at large enough magnetic field $B_0$ ($\gamma_iB_0\gg A_{i,zx},A_{i,zy}$). We can furthermore ignore the constant energy shift from the quadrupole moment for our spin-1/2 nuclear spins. We approximate the electronic spin as a two-level system consisting of the $m_s=0$ ($|\uparrow\rangle$) and $m_s=-1$ ($|\downarrow\rangle$) magnetic levels. For the qubit register we define the computational subspace for the configuration where the electron is in the $|\downarrow\rangle$ state. The computational basis states are then given by the eigenstates of the $I_{i,z}$ nuclear spin operators.
Under these approximations the static Hamiltonian is 
\begin{eqnarray}
	H_{\mathrm{static}}=\frac{\omega_{el}}{2}\sigma_z + 
	\sum_i  \left(- \gamma_i B_0 - \frac{A_{i,zz}}2  + \frac{A_{i,zz}}2 \sigma_z \right)I_{i,z}\,,
\end{eqnarray}
where $\omega_{el} =  D - \gamma_{NV}B_0 $ and $\sigma_z$ is the Pauli-z matrix. We now consider the time-dependent drive terms in the frame rotating with $H_{\mathrm{static}}$.

We first consider the driving of the nuclear spins, where $B_1(t)$ is in the radio frequency regime. 
\begin{eqnarray}
	\hat{H}_{SQ}(t)=-\exp(i H_{\mathrm{static}}t)\sum_i \gamma_i B_1(t) I_{i,x} \exp(-i H_{\mathrm{static}}t)\\
	= -|\downarrow\rangle\langle \downarrow |\sum_i \left[ I_{i,x} \cos (\omega_i t) + I_{i,y} \sin (\omega_i t) \right]\gamma_i B_1(t)\nonumber\\
	-|\uparrow\rangle\langle \uparrow |\sum_i \left[ I_{i,x} \cos (\gamma_i B_0 t) + I_{i,y} \sin (\gamma_i B_0 t) \right] \gamma_i B_1(t)\,,
\end{eqnarray}
where $\omega_i=\gamma_i B_0+A_{i,zz}$.
We want to apply pulses on the nuclear spins only when the electron is (to high precision) in $|\downarrow\rangle$ and thus make the further simplification
\begin{equation}
	\label{eq:single_qubit_ham}
	\hat{H}_{SQ}(t) \approx - \sum_i \left[ I_{i,x} \cos (\omega_i t) + I_{i,y} \sin (\omega_i t) \right] \gamma_i B_1(t).
\end{equation}
Since the nuclear-spin transition frequencies $\omega_i$ are different for every spin ($\omega_i\neq \omega_j$ for $i\neq j$), spectrally narrow pulses can address the spins individually~\cite{Wrachtrup2006,Chen2020}.

Second, we consider the driving of only the electron spin with $B_1(t)$ in the microwave regime. We obtain
\begin{eqnarray}
	\hat{H}_{NV} &=& \exp(i H_{\mathrm{static}}t)\frac{\gamma_{NV}}{\sqrt 2} B_1(t) \sigma_x\exp(-i H_{\mathrm{static}}t)\\
	&=&\large\frac{\gamma_{NV}}{\sqrt 2} B_1(t) \sum_{l=1}^{2^N}  \big[
		 \cos \left(\omega_{el} t+\sum_{j=1}^N F(s_{lj})\beta_j t\right)\sigma _ { x }\nonumber\\
		&&- \sin \left(\omega_{el} t+\sum_{j=1}^N F(s_{lj}) \beta_j t\right) \sigma_y\big]\otimes \ket{l}\!\bra{l}  \nonumber\\
		&=&\large\frac{\gamma_{NV}}{\sqrt 2} B_1(t) \sum_{l=1}^{2^N}  \big[
		\cos \left(\omega_{el} t+\tilde\omega_l t\right)\sigma _ { x }\nonumber \\
		&&- \sin \left(\omega_{el} t+\tilde\omega_l t\right) \sigma_y \big]\otimes \ket{l}\!\bra{l}  \,,\label{eq:multi_qubit_ham}
\end{eqnarray}
where $\beta_i=A_{i,zz}/2$ denotes half of the hyperfine splitting for each qubit,
$s_{lj}  \in \{0,1\}$ are the binary spin states of the $j^{th}$ computational qubit of $|l \rangle$, and $|l \rangle$ denotes the $l^{th}$ computational basis state in the $2^N$ dimensional Hilbert space. For example, $|1\rangle=|0\dots00\rangle$, $|2\rangle=|0\dots01\rangle$, and so on. $F(1) = +1, F(0) = -1$ is the parity function which determines with which sign the hyperfine splitting $\beta_j$ enters the level energy $\tilde\omega_l=\sum_{j=1}^N F(s_{lj})\beta_j$ of the levels $| l \rangle$. The input value $ s_{lj} $ to $F(s_{lj})$ is exactly the computational state of the $j^{th}$ qubit of state $|l\rangle$. E.g., for $|000\rangle$, all inputs would be 0.
If we explicitly write out the Hamiltonian for two physical qubits, we obtain
\begin{eqnarray}
\hat{H}_{NV}(t) = \frac{\gamma_{NV}}{\sqrt 2} B(t) \sum_{\alpha,\beta = 0}^1 [\cos (\omega^{\alpha\beta}_{ij} t) \sigma_x - \sin (\omega^{\alpha\beta}_{ij} t )\sigma_y] \nonumber \\ \qquad\qquad \otimes \ket{\alpha\beta}\!\bra{\alpha\beta},
\end{eqnarray}
where $i$ and $j$ label a pair of two distinct nuclei coupled to the NV center, $\ket{\alpha\beta}\!\bra{\alpha\beta}$ denotes the projection operator onto computational states spanned by the pair of nuclei, and $\omega^{\alpha\beta}_{ij}$ denotes the rotation frequency corresponding to the $i,j$ nuclei pair and the $\ket{\alpha\beta}\!\bra{\alpha\beta}$ projection operator as given in Table~\ref{tab:Freq_dict}. Note that we can find a similar form of the Hamiltonian for group-IV defects like silicon, tin or germanium vacancy centers that have recently gained interest due to their spin-photon coupling properties as candidates for quantum repeater or optically-scalable quantum processing nodes~\cite{ruf2021quantum,wang2024transform, bradac2019quantum, stas2022robust, sukachev2017silicon, Karapatzakis2024, Grimm2025}. The key difference in the model in this context is that the electron spin is a two-level system and the hyperfine interaction enters the Hamiltonian in a slightly different way (see appendix~\ref{app:groupiv}).

\begin{table}[h]
	\caption{Rotation frequencies corresponding to each computational state projector for nuclei pair $i,j$ for an example of two nuclear spins.}
	\centering
	\begin{tabular}{|c|c|}
		\hline
		$\bm{\ket{\alpha\beta}\!\bra{\alpha\beta}}$ & $\bm{\omega^{\alpha\beta}_{ij}}$ \\
		\hline
		$\ket{00}\!\bra{00}$ & $\omega_{el} - \beta_i - \beta_j$ \\
		$\ket{01}\!\bra{01}$ & $\omega_{el} - \beta_i + \beta_j$ \\
		$\ket{10}\!\bra{10}$ & $\omega_{el} + \beta_i - \beta_j$ \\
		$\ket{11}\!\bra{11}$ & $\omega_{el} + \beta_i + \beta_j$ \\
		\hline
	\end{tabular}
	\label{tab:Freq_dict}
\end{table}

If we drive the microwave in such a way that the electron spin starts in $|\downarrow\rangle$ and ends up in $|\downarrow\rangle$ again, each basis state $|l\rangle$ accumulates a different phase, while the electron spin is restored to its initial state.
Thus, the electron spin stays decoupled from the nuclear spin states between any two such operations and merely mediates entangling operations. This has the additional advantage that dephasing plays a role only during the fast microwave gates. Since the  microwave also does not flip the nuclear spins (due to the different frequency regime), the effective gate operation is thus a multi-qubit phase gate on the $N$ nuclear spins of the qubit register.

The total dynamics of the system in the rotating frame is thus determined by the unitary evolution 
$$ U(T,0) = \mathcal{T}\exp \left \{ - i \int^T_0 \left [ \hat{H}_{SQ}(t) + \hat{H}_{NV}(t)\right ] dt \right \}, $$
where $\mathcal{T}$ is indicating time-ordering of the exponential and where the choice of the control field allows us to activate the single-qubit Hamiltonian, Eq.~\eqref{eq:single_qubit_ham}, or the multi-qubit Hamiltonian, Eq.~\eqref{eq:multi_qubit_ham}.
In this way we have all the ingredients to design arbitrary single-qubit gates and certain multi-qubit phase gates, as building blocks for universal quantum computing. The pulse generation for these gates is discussed in detail in Section \ref{section:pulse_gen}. Now, we first want to show how we can decompose an arbitrary quantum circuit into the elementary, or native, gates of our system.

\section{Circuit design via algebraic decomposition}
\label{section:cartan}
In this section we show how we can make use of algebraic decomposition to decompose an arbitrary quantum circuit (given by its final unitary) into the native gates of our system. We start by reviewing the Cartan and $KAK$ decomposition and then show how to apply it to our scenario.
The Cartan decomposition~\cite{houdayer2013type,Khaneja2000,khaneja2001cartan,bullock2004} is a set of theorems applicable to semi-simple Lie algebras, of which $SU(2^{N})$ is an example. Closely related to the decomposition method is the $KAK$ theorem, which decomposes a unitary into local operations and interaction terms~\cite{Khaneja2000,Wieser2013}.

\begin{definition}
	Let $\mathfrak{g}$ be a semi-simple Lie algebra. There exists a set of elements $\mathfrak{m}$ and a sub-algebra $\mathfrak{k}$ where $\mathfrak{m}, \mathfrak{k} \subseteq \mathfrak{g}$, such that
	$$ \mathfrak{g} = \mathfrak{k} \oplus \mathfrak{m}$$
	$$[\mathfrak{k}, \mathfrak{k}] \subset \mathfrak{k}, [\mathfrak{m},\mathfrak{k}] \subset \mathfrak{m}, [\mathfrak{m}, \mathfrak{m}] \subset \mathfrak{k}.$$
	The pair  $\mathfrak{k}$ and  $\mathfrak{m}$ is called a Cartan pair and can be generated by applying a chosen involution $\theta$ on the Lie algebra $\mathfrak{g}$. 
	$$\theta: \mathfrak{g} \mapsto \mathfrak{g}$$
	An example of such involution is
	$$\theta (k \in \mathfrak{k}) = k, \theta (m \in \mathfrak{m}) = -m$$
	One can easily verify that the resulting sets follow the above commutation relations.
\end{definition}
Note that it is exactly the freedom of choice in the involution that allows for different decomposition regimes tailored to different system Hamiltonians. For example, it can be used to bring the resulting generator $m$ in the form of $\sigma_x$ or $\sigma_z$ operators, as will be shown later.

\begin{definition}
	Given a Cartan pair, the Cartan subalgebra $\mathfrak{h}$ is defined to be the maximal Abelian subalgebra residing in $\mathfrak{m}$.
\end{definition}

\begin{theorem}
	Given a unitary $ U \in SU(2^{N})$, $\exists$ a decomposition 
	$$ U = K_1  \exp (Z_1)  K_2  \exp(Y)  K_3  \exp (Z_2)  K_4, $$
	where $K_1, K_2, K_3, K_4 \in SU(2^{N-1} \otimes U(1)), Y \in \mathfrak{h}$, Cartan subalgebra of the pair $(\mathfrak{su}\small{(2^N)},\,\mathfrak{su_k}\small{(2^N)})$ and $Z_1, Z_2 \in \mathfrak{f},$ a Cartan subalgebra of the pair
	$(\overline{ \mathfrak{su_k}\smash\small{(2^{N})}}$, $\mathfrak{su_{k0}}\small{(2^N)} )$,
where $\overline{\mathfrak{su_{k}}\small{(2^N)}}=\text{span}\{ O\otimes \id , O \otimes \sigma_z| O\in  \mathfrak{su}(\small{2^{N-1}})\}$ and $\mathfrak{su_{k0}}\small{(2^N)}=
\text{span}\{ O\otimes \id | O\in  \mathfrak{su}(\small{2^{N-1}})\}$.
\end{theorem}
\noindent For more  details, see Ref.~\cite{Khaneja2000,Wieser2013}.

The decomposition protocol involves three steps. First, the native Lie algebra elements in the system Hamiltonian are identified that correspond to the native gates of the system. Second, a recursive Cartan decomposition is designed, with the choice of Cartan involutions that produce a decomposed sequence consisting of only unitaries that are generated by the native Lie algebra elements of the Hamiltonian. In the final step, a heuristic procedure can help to shorten the circuit in some scenarios. This is described in the following sections.

\subsection{Lie algebra structure of Hamiltonian}
To identify the native Lie algebra elements present in the system Hamiltonian, we start by examining the single-qubit Hamiltonian given in Eq.~\eqref{eq:single_qubit_ham}, which is addressable by radio frequency control pulses. It is easy to see that the Lie algebra elements for the single-qubit operations are simply
$$i \text{span}\left \{ \sigma_{i,x},\,\sigma_{i,y} \right \},$$
with the Pauli-x $\sigma_{i,x}$ and Pauli-y $\sigma_{i,y}$ matrices on qubit $i$.
With these matrices we can generate the local rotations $R_x(\zeta) = \exp\big(\frac{i \zeta}{2} \sigma_x\big)$ and $R_y(\zeta) = \exp\big(\frac{i \zeta}{2} \sigma_y\big)$ directly, and $R_z(\zeta) = R_y(\pi/2)R_x(\zeta) R_y(-\pi/2)$ indirectly.
Instead, the multi-qubit Hamiltonian, Eq.~\eqref{eq:multi_qubit_ham}, allows us to generate the diagonal elements of the nuclear $2^N \times 2^N$ unitary matrix, namely
$$\mathcal{S}_N = \left\{
\begin{bmatrix}
	e^{i\phi_1} & & \\
	& \ddots & \\
	& & e^{i\phi_{2^{{}^N}}} \\
\end{bmatrix}\Bigg| \phi_i \in \mathbb{R} (i=1,\dots 2^N) \right\}\,.$$
How exactly these gates can be generated will be discussed in more detail in Section~\ref{section:gate_design}.

\subsection{Decomposition algorithm}\label{sec:decomposition_algorithm}
Now, we are fully equipped for the recursive decomposition algorithm of a given unitary target gate $ G \in SU(2^N)$. We start from an idea presented in Ref.~\cite{Nakajima2003} and tailor it to our case by adding an additional layer in the decomposition algorithm that allows us to obtain only the local gates and the diagonal multi-qubit gates described above. In each layer, an involution is chosen to decompose the gates into elements of the Cartan pair [$\mathfrak{k}, \mathfrak{m}$] defined by the involution.

\begin{enumerate}
	\item We start by the involution
	$$\theta (x) = \sigma_{N,z}\,x\,\sigma_{N,z},$$ where $\sigma_{N,z} = \id_{N-1} \otimes \sigma_z$ 
	and the Cartan pair ($\mathfrak{k}, \mathfrak{m}$) is:
		$$
	\mathfrak{m} = \text{span}\left \{(i \id_{N-1} \oplus \mathfrak{su}(2^{N-1}))\otimes \sigma_x,\sigma_y \right\}
	$$
	\begin{eqnarray}
			\mathfrak{k} = \text{span} \left \{(i \id_{N-1}\!\oplus \mathfrak{su}(2^{N-1}))\otimes \sigma_z\right \}\nonumber\\
			\cup\; \text{span}\left \{\mathfrak{su}(2^{N-1}) \!\otimes \id\right \}\,.\nonumber
	\end{eqnarray}
	\item For the decomposition $G = \exp(k) \exp(m)$, we can calculate $m^2 = \theta(g^{\dagger})g$, where $g$ is the Lie algebra elment corresponding to $G=\exp(g)$.
	\item We can rotate $m$ into the corresponding Cartan subalgebra element $h$, such that $m = K_1 h K_1^{\dagger}$. Since the maximal Cartan subalgebra for a given Cartan pair is not unique, we can choose it such that the decomposed sequence consists of only nuclear rotations and diagonal entangling gates, namely
	$$\mathfrak{h} = i \text{span} \left \{\ket{j}\!\bra{j} \otimes \sigma_x \right \}.$$
	An algorithm for computing $h$ for our choice of involution and $\mathfrak{h}$ can be found in Section 3.2.1 of Ref.~\cite{Nakajima2003}. Following the reference, we obtain
	$$ \exp(h) = \sum^{2^{N-1}}_{j = 1} \ket{j}\!\bra{j} \otimes R_x (\zeta_j) ,$$
	and with an additional local $y$-rotation we can bring it to diagonal form: $R_z(\zeta_j) = R_y(\pi/2)R_x(\zeta_j) R_y(-\pi/2)$.
	\item Now we have $ G = K K_1 H K_1^\dagger$, $H = 
	\exp(h)$, where only $H$ is of the desired form. We are thus left with $K K_1$  and $K_1^\dagger$ that belong to $\exp(\mathfrak{k})$, which we now further decompose. In contrast to the involution in step 1., we choose the following involution instead
	$$\theta_1 (x) = \sigma_{N,x}\,x\,\sigma_{N,x}, $$ which produces the following Cartan pair
	$$
	\mathfrak{m}_1 = \text{span}\left \{(i \id_{N-1} \oplus \mathfrak{su}(2^{N-1}))\otimes \sigma_z\right\}
	$$
	$$
	\mathfrak{k}_1 = \text{span}\left \{\mathfrak{su}(2^{N-1}) \otimes \id\right \}.
	$$
	with the corresponding maximal Abelian Cartan subalgebra
	$$\mathfrak{h}_1 = i\text{span} \left \{\vert j \rangle \langle j \vert \otimes \sigma_z \right \},$$ which is intrinsically diagonal.
	\item At this level of decomposition, we utilize an alternative method that directly computes the corresponding $m_l, m_r\in \mathfrak{m}_1$ for $K_l=K K_1$  and $K_r=K_1^\dagger$ and the unique choice of its projective components in the $\mathfrak{k}_1$ subalgebra. Then, we can achieve the further decomposition of $m_{l,r}$ into $h_{l,r}$ and $k_{l,r}$.
	We thus calculate
	$$K K_1 = \exp(k_l) \exp(m_l)$$
	$$m_l = u_l \,\tilde M\, v_l \otimes \sigma_z = (u_l \otimes \mathbbm{1})\,h_l\,(v_l \otimes \mathbbm{1}) $$
	where $h_l = \tilde M \otimes \sigma_z$ and $u_l, v_l \in \mathfrak{su}(2^{N-1}).$ Similarly, the procedure follows for $K_1^{\dagger} = \exp(k_r)U_r\exp(h_r) V_r $.
	The details of this step of the algorithm can be found in appendix~\ref{appendix: A}.
	\item We have now  arrived at the intermediate result in form of Theorem 1. 
	$$G = K_l U_l \exp(h_l)V_l \exp(h) K_r U_r\exp(h_r)V_r\,.$$
    The elements  $h_l$, $h$ and $h_r$ have the desired form. All other elements ($K_l$, $U_l$, $V_l$, $K_r$, $U_r$ and $V_r$) are effectively elements of $SU(2^{N-1})$ and can thus be decomposed by restarting from step 1. of the decomposition algorithm.
\end{enumerate}

\subsubsection{Variant for Two-qubit Gates}\label{sec:variant_two_qubit}
Note that for the last iteration (decomposition of SU(4)) there exist numerical algorithms~\cite{Earp2005} that efficiently realize the Cartan decomposition with the canonical choice of the $\mathfrak{su}(4)$ Lie algebra partition, \textit{i.e.} $$\mathfrak{h} =i\text{span} \{ \sigma_x \sigma_x , \sigma_y\sigma_y,\sigma_z \sigma_z \}$$
$$ \mathfrak{m} =i\text{span} \{  \sigma_x \sigma_y , \sigma_x \sigma_z, \sigma_y \sigma_x , \sigma_y \sigma_z, \sigma_z \sigma_x, \sigma_z \sigma_y\} \oplus \mathfrak{h}$$
$$ \mathfrak{k} = i\text{span}\{  \sigma_x \mathbbm{1}, \mathbbm{1} \sigma_x , \sigma_y \mathbbm{1}, \mathbbm{1} \sigma_y , \sigma_z \mathbbm{1},\mathbbm{1} \sigma_z \} $$
$$ \mathfrak{su}(4) = \mathfrak{k} \oplus \mathfrak{m}\,.$$
This alternative sometimes has the benefit of producing decomposed elements with coefficients that are integer fractions of $\pi$. We will point out in the remainder of the text when we will use this alternative.

\subsection{Decomposition Heuristics}
\label{section:heur}
As the complexity of recursive Cartan Decomposition scales exponentially with the number of qubits involved (in each level of the recursive decomposition all gates obtained by the previous level of the decomposition potentially have to be decomposed again), in general, to decompose an N-qubit unitary in $SU(2^N)$, about $\mathcal{O}(4^{N-1})$ single-qubit gates and entangling gates are required. The performance of the algorithm thus deteriorates if the qubits in the circuit are only "loosely" connected, and an example of such a loosely connected circuit is the Quantum Fourier Transform. In such cases, heuristic measures are taken to reduce the dimension of the problem such that the N-qubit unitary reduces to a product of unitaries in smaller Hilbert spaces $$U \in SU(2^N) = \Pi_i U^i$$ where each of the unitaries in the product has a dimension smaller than $N$, $$ U^i \in SU(2^{N_i}),$$ such that $\Sigma_{i}N_i = N$. To this end, three main steps are employed:
\begin{enumerate}
	\item \emph{Direct reduction}: There are non-participating qubits whose corresponding Hilbert space can be traced over to obtain an equivalent unitary $\underset{\{q_{np}\}}{\Tr}U \in SU(2^{N - |q_{np}|})$ involving only $ N - |q_{np}| < N$ qubits, where $\{q_{np}\}$ is the set of non-participating qubits. 
	\item \emph{Product Unitary}: The algorithm detects at every level of decomposition if the target unitary is a direct product of two or more unitaries spanning disjoint sets of the qubits. For example, all disjoint sets partitioning a three-qubit system are [\{1\}, \{2\}, \{3\}], [\{1,2\},\{3\}], [\{1,3\},\{2\}] and [\{1\},\{2,3\}]; additionally, there is the trivial case of [\{1,2,3\}] which represents that no true partition is found. Denoting the set of all  non-trivial partitions as $\mathcal{S}$, the algorithm returns one or more elements $\mathfrak{p} \in \mathcal{S}$, such that $U \in SU(2^N) = \Pi_\mathfrak{p} U_p$. Since any such solution isn't trivial, the product terms naturally have dimensions less than $N$.
	\item \emph{Known circuit elements}: Given that a possible usage of this algorithm is partial or complete recompilation of a known circuit such as QFT, one method to surpass the sub-optimal performance when qubits are loosely connected is to "unwrap" known operations from the unitary as follows: A set of common single-qubit and two-qubit gates is selected. Again, at every level of decomposition, the algorithm checks if left or right multiplication of the target unitary by any of the gates reduces the dimension of the unitary, i.e.:
	\begin{eqnarray}
	\exists U_t \in \{\text{Basic gate set}\}, s.t. \nonumber\\U_t U = U' ,
	U'  \in SU(2^{N'} ), N' < N \nonumber\\
	\text{or}\, U U_t  = U' , U'  \in SU(2^{N'}), N'  < N.
	\end{eqnarray}
\end{enumerate}
The basic gates we check include single-qubit gates and two-qubit gates. The single-qubit gates are rotations around $\sigma_x$ and $\sigma_y$ with a rotation angle $\pi/2^k$ ($k=0,\dots N$) on any qubit. The two-qubit gates are CNOT, SWAP and controlled phase gates with a rotation angle $\pi/2^k$ ($k=0,\dots N$). The two-qubit gates are checked between each pair of qubits.

\section{Gate Design via Pulse Shaping}\label{section:gate_design}
In this section we show how we can generate the necessary gates for the decomposition of Sec.~\ref{section:cartan} with the system described in Sec.~\ref{section:NV_section}. We start by defining the gate fidelity and then discuss the generation of single-qubit and entangling gates separately.

\subsection{Gate Fidelity}
When choosing an appropriate definition of the fidelity, one needs to keep in mind that the computational space is just a subsystem of the larger dynamical system. Indeed, the full time propagator $U_f$ involves the electron spin, which is only an ancilla qubit that should return to its initial $\vert \downarrow \rangle$ state after each gate operation throughout the protocol. This is equivalent to projecting the whole system onto the $\vert \downarrow \rangle\langle \downarrow \vert \otimes \id_N$ subspace, where $\id_N$ is the identity operator over the $N$ logical (nuclear spin) qubits.  

In this way we can calculate the reduced time-propagator $U_r=\Tr_{NV}[(\vert \downarrow \rangle\langle \downarrow \vert \otimes \id_N) U_F (\vert \downarrow \rangle\langle \downarrow \vert \otimes \id_N)],$
where $\Tr_{NV}$ denotes taking the trace over the first ancilla electron qubit.
A target gate, given by the perfect unitary $U_p \in SU(2^N)$ is thus compared to not the full time-propagator involving $N+1$ qubits, $U_f \in SU(2^{N+1})$, but the reduced time-propagator $U_r$.
We can thus define the gate fidelity as 
$$\mathfrak{F} = \vert \Tr(U_r^{\dagger}U_p)/\Tr(U_p^{\dagger} U_p)\vert,$$
where $\Tr(U_p^{\dagger}U_p) = 2^N$ serves as a normalization factor.

\subsection{Single-Qubit Gates}
\label{section:pulse_gen}
Due to the large separation in transition frequencies, we calculate the control pulses $B_1(t)$ for the single-qubit gates by only considering the single-qubit Hamiltonian Eq.~\eqref{eq:single_qubit_ham}. For the derivation of the single-qubit pulses, we furthermore assume that we can spectrally resolve the single nuclear spins. Later in the numerical evaluation of the fidelity we will consider the full Hamiltonian without this assumption.
We start by considering a pulse that should resonantly drive spin $j$ with constant amplitude $B_c$ and a control phase $\phi_c$:
\begin{equation}
B_1(t) = B_c \cos(\omega_j t+\phi_c).
\end{equation}
The Hamiltonian  Eq.~\eqref{eq:single_qubit_ham} then takes the form
\begin{eqnarray}
	H_{SQ}=-(I_{j,x} \cos \omega_j t + I_{j,y} \sin \omega_j t ) \gamma_j B_c \cos(\omega_j t+\phi_c)\\
	= - \frac{\gamma_j B_c}{2}\begin{pmatrix}
		0 & e^{- i \omega_j t}\\ e^{i \omega_j t} &0
	\end{pmatrix} \frac{e^{ i \omega_j t + i\phi_c}-e^{- i \omega_j t-i\phi_c}}{2}\\
	\approx - \frac{\gamma_j B_c}{4}\begin{pmatrix}
		0 & e^{+i\phi_c}\\ e^{- i \phi_c} &0
	\end{pmatrix} \\
	=- \frac{\gamma_j B_c}{4}\left(\cos\phi_c \sigma_x - \sin\phi_c \sigma_y\right)\,
\end{eqnarray}
where the approximation consists  in the rotating wave approximation (RWA) of neglecting the terms oscillating with the double frequency $2\omega_j$ and is justified for $\omega_j\gg \gamma_j B_c$, i.e., for the typically slow radio frequency pulses. If we irradiate this pulse at constant amplitude for some time $T$, we thus obtain a single-qubit gate on qubit $j$ of the form
$$\exp(i \zeta (n_x \sigma_x + n_y \sigma_y)),$$
where the rotation angle $\zeta=- \frac{\gamma_j B_c T}{4}$ and the rotation axis (given by $n_x=\cos\phi_c$ and $n_y=\sin\phi_c$) can be chosen arbitrarily by the control parameters $B_c$ and $\phi_c$.
In the general case of a rotation about an arbitrary axis in the Bloch sphere, one can readily convert the general arbitrary rotation $\exp (i \mathcal{\zeta} \hat{n} \cdot \overrightarrow{\sigma})$ by XYX decomposition, \textit{i.e.} 
$$\exp (i \mathcal{\zeta} \hat{n} \cdot \overrightarrow{\sigma}) = \exp (i \zeta_1 \sigma_x) \exp (i \zeta_2 \sigma_y) \exp (i \zeta_3 \sigma_x),$$
for some real coefficients $\zeta_1$, $\zeta_2$, $\zeta_3$.

\subsection{Entangling Phase Gates}
For the multi-qubit entangling gates we drive the multi-qubit Hamiltonian Eq.~\eqref{eq:multi_qubit_ham}
at the resonance frequency $\omega_{el}$ of the bare electron transition
\begin{eqnarray}
	B_1(t)=\tilde{B}(t)\cos(\omega_{el} t)\,,
\end{eqnarray}
where $\tilde{B}(t)$ is an envelope function that allows us to control the dynamics. If we perform a RWA similarly to the single-qubit case, we obtain
\begin{eqnarray}
	H_{NV}(t)&=&\frac{\gamma_{NV}}{\sqrt 2} \tilde{B}(t)\cos(\omega_{el} t) \sum_{l=1}^{2^N}  \big[
	\cos \left(\omega_{el} t+\tilde\omega_l t\right)\sigma _ { x } \nonumber\\
	  &&- \sin \left(\omega_{el} t+\tilde\omega_l t\right) \sigma_y\big]\otimes \ket{l}\!\bra{l} \\
	 &\approx& \frac{\gamma_{NV}}{2 \sqrt 2} \tilde{B}(t) \sum_{l=1}^{2^N}  \big[
	 \cos \left(\tilde\omega_l t\right)\sigma _ { x } \nonumber\\
	 &&- \sin \left(\tilde\omega_l t\right) \sigma_y\big]\otimes \ket{l}\!\bra{l}\,.
\end{eqnarray}
Note that we cannot selectively drive the different contributions given by the interaction frequencies $\tilde\omega_l$ because we want to achieve fast gates below the dephasing limit. We therefore cannot make further simplifications but have to solve the time dynamics numerically to optimize the control shape $\tilde{B}(t)$. We start by discretizing the time evolution with $N_t$ equidistant time points $t_i=i \tau$, with a time step $\tau$, and introducing the shorthand notation $f_l(t)=\frac{\gamma_{NV}}{2 \sqrt 2} \tilde{B}(t) \cos \left(\tilde\omega_l t\right)$ and $g_l(t)=\frac{\gamma_{NV}}{2 \sqrt 2} \tilde{B}(t) \sin \left(\tilde\omega_l t\right)$. We obtain:
\begin{align*}
	& U(T,0)=\mathcal{T}\exp \left(-i \int_0^T\hat{H}_{NV} (t)dt\right) \\
	& =\prod_{i=1}^{N_t} \exp \left(-i\sum_{l=1}^{2^N}\left[f_l\left(t_i\right) \sigma_x - g_l\left(t_i\right) \sigma_y \right]\otimes|l \rangle\langle l |\tau\right) \\
	& =\prod_{i=0}^{N t} \sum_{l=1}^{2^N} \exp \left(-i\left(f_l\left(t_i\right) \sigma_x-g_l\left(t_i\right) \sigma_y\right) \tau \right)\otimes|l \rangle\langle l | \\
	& =\prod_{i=1}^{N_t} \sum_{l=1}^{2^N} \exp \left(-i \frac{\zeta_{il}}{2}  \hat{R}_{il} \cdot \vec{\sigma} \right) \otimes|l \rangle\langle l |\\
	& =\sum_{l=1}^{2^N}  \left(\prod_{i=1}^{N_t} \exp \left(-i \frac{\zeta_{il}}{2}\hat{R}_{il} \cdot \vec{\sigma} \right) \right) \otimes|l \rangle\langle l |\,,
\end{align*}
where $\zeta_{il}$ is the rotation angle and $R_{il}$ the normalized rotation vector such that $\frac{\zeta_{il}}{2}  \hat{R}_{il} \cdot \vec{\sigma} = \left(f_l\left(t_i\right) \sigma_x-g_l\left(t_i\right) \sigma_y\right) \tau$.
Now, the time-evolution of the system reduces to the calculation of the products
$$\prod_{i=1}^{N_t} \exp \left(-i \frac{\zeta_{il}}{2}\hat{R}_{il} \cdot \vec{\sigma} \right)= \exp \left(-i \frac{\zeta_{l}}{2}\hat{R}_{l} \cdot \vec{\sigma} \right) $$
which can be done efficiently via the rotation addition formula~\cite{NC}:

\begin{theorem}
	A rotation by an angle $\vartheta_1$ about an axis $\hat{n}_1$ followed by another rotation by an angle $\vartheta_2$ about an axis $\hat{n}_2$, denoted by $\exp (i \vartheta_2 / 2 \hat{n}_2 \cdot \vec{\sigma})\exp (i \vartheta_1 / 2 \hat{n}_1 \cdot \vec{\sigma})$, is equivalent to an overall rotation by an angle $\vartheta_{12}$ about an axis $\hat{n}_{12}$, where $\vartheta_{12}$ and  $\hat{n}_{12}$ are given by
	$$
	\begin{array}{rcl}
		c_{12}&=&c_1 c_2-s_1 s_2 \hat{n}_1 \cdot \hat{n}_2 \\
		s_{12} \hat{n}_{12}&=&s_1 c_2 \hat{n}_1+c_1 s_2 \hat{n}_2-s_1 s_2 \hat{n}_2 \times \hat{n}_1,
	\end{array}
	$$
	where $c_i=\cos \left(\vartheta_i / 2\right), s_i=\sin \left(\vartheta_{i / 2}\right)$.
\end{theorem}

The target gate now corresponds to $\zeta_l\hat{R}_l\cdot \hat{z} =\phi_l$ (i.e., the z-component of the rotation has to yield the desired phase on the state $|l\rangle$) and $\zeta_l\hat{R}_l\cdot \hat{x} =\zeta_l\hat{R}_l\cdot \hat{y} =0$ (i.e., the electron spin has to return to the initial state $|\downarrow\rangle$ after the gate).
To finally optimize the diagonal entangling gates $\mathcal{S}_N$ we combine these three conditions in the figure of merit (FoM)
\begin{equation}\label{eq:FoM_entangling_gate}
	\mathrm{FoM}=\prod_{l=1}^{2^N}\cos (\zeta_l\hat{R}_l\cdot \hat{z} -\phi_l) \cos (\zeta_l\hat{R}_l\cdot \hat{x} ) \cos (\zeta_l\hat{R}_l\cdot \hat{y} )
\end{equation}
and inspired by the dressed chopped random basis (dCRAB) algorithm~\cite{Caneva2011,Rach2015, crab} for QOC we adopt the following ansatz for the control pulse:
\begin{equation}
\tilde B(t) = \Sigma_{i=1}^{N_{freq}} a_i \cos(\Omega_i t + \varphi_i) + b_i\sin(\Omega_i t + \psi_i)\,,
\end{equation}
where $\Omega_i$ are harmonic Fourier frequencies that shape the envelope of the control pulse; we choose $\Omega_i = 2i\pi/T$, where $T$ is the gate time. Note that one could also choose these frequencies randomly~\cite{crab}. We choose $N_{freq}=10$ and optimize the coefficients $a_i$ and $b_i$ as well as the phases $\varphi_i$ and $\psi_i$ (within the interval $[-\pi,\pi]$) by maximizing the FoM via the L-BFGS algorithm in its Julia Optim version.

\section{Numerical Implementation}\label{sec:numerical_implementation}
\begin{table*}
	\caption{Table with physical parameters used for the numerical simulation~\cite{Su2014,Felton2009}. The two carbon spins correspond to the L-family ($^{13}$C$_1$) and S-family ($^{13}$C$_2$).}
	\centering
	\begin{tabular}{|c|c|c|}
		\hline
		Parameter & Label & Value ($2\pi$)\\
		\hline
		Zero Field Splitting of NV & $D$ & 2.87 GHz/T\\
		Gyromagnetic ratio of NV & $\gamma_{NV}$ & 28.024 GHz/T\\
		Gyromagnetic ratio of $~^{15}$N & $\gamma_{~^{15}N}$ & -4.32 MHz/T\\
		Gyromagnetic ratio of $~^{13}$C$$ & $\gamma_{~^{13}C}$ & 10.708 MHz/T\\
		Hyperfine structure constant of $~^{15}$N & $\beta_{~^{15}N}$ & 1.515 MHz\\
		Hyperfine structure constant of $~^{13}$C$_1$& $\beta_{~^{13}C}$ & 0.49 MHz \\
		Hyperfine structure constant of $~^{13}$C$_2$& $\beta_{~^{13}C}$ & 0.206 MHz \\
		\hline
	\end{tabular}
	\label{tab:system_params}
\end{table*}
The circuit decomposition algorithm of Sec.~\ref{section:cartan} has been implemented in Julia. For the time evolution of the individual gates, we numerically calculate the time evolution by integrating the Schrödinger equation for the full Hamiltonian $H(t)=H_{SQ}(t)+ H_{NV}(t)$ with $N_t$ time steps with $\tau=1\,$ns:
\begin{eqnarray}
	U(t)=\prod_{j}^{N_t} \exp(-i H(t_j) \tau)\,.
\end{eqnarray}
The parameters that were used in the simulations are given in Table~\ref{tab:system_params}. The table shows only the parallel hyperfine splitting $\beta_i=A_{i,zz}/2$. The nitrogen spin has vanishing perpendicular hyperfine splitting ($A_{zx}=A_{zy}=0$). The two carbon spins correspond to the L-family ($^{13}$C$_1$) and S-family ($^{13}$C$_2$) (see Ref.~\cite{Su2014}) and have a perpendicular hyperfine splitting that is one order of magnitude smaller than the parallel hyperfine splitting and we neglect it in the simulations.

\section{Results}
\label{section:results}
In this section, we will demonstrate the decomposition algorithm and the pulse generation for several common problems with increasing complexity. We will first show how to generate a CNOT gate and a SWAP gate between two nuclear spin qubits ($^{15}$N and $^{13}$C$_1$). We will then derive a control pulse sequence for QFT for three qubits ($^{15}$N, $^{13}$C$_1$ and $^{13}$C$_2$). For each of these examples we will first decompose the target unitary into a sequence of gates out of our native gate set, calculate the control pulses for each gate in the circuit and then calculate the time evolution for the circuit and evaluate the fidelity with respect to the target unitary. Finally, we will show how the decomposition algorithm performs for random circuits for up to six qubits.

\begin{figure}
	\centering
	\includegraphics[width = 0.45\textwidth]{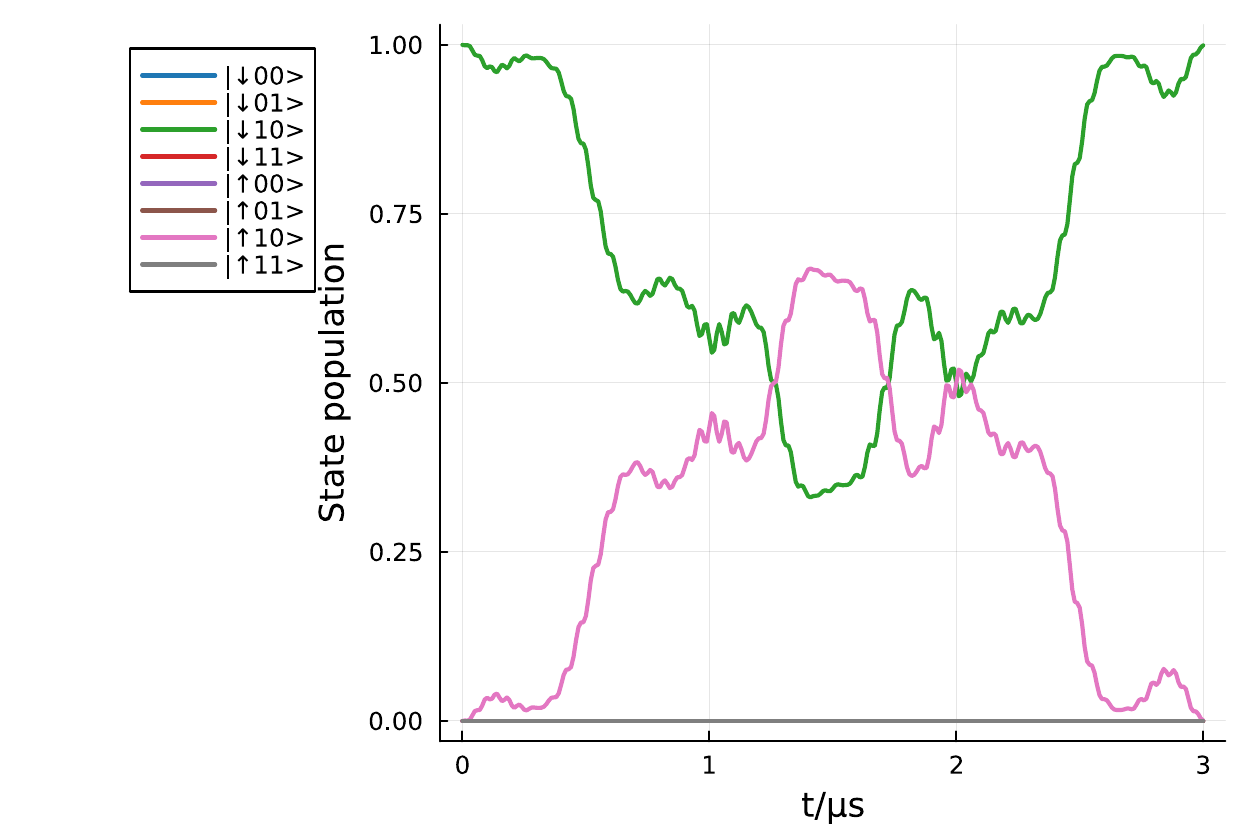}
	\caption{Time dynamics of the $ZZ$ gate. We show the population of the basis states for the initial state $|\downarrow 10\rangle$. We can see that the electron is (partially) excited to $|\uparrow\rangle$ so that the nuclear basis states can accumulate the entangling phases ($\pi/4$, $-\pi/4$, $-\pi/4$, and $\pi/4$, respectively) and finally the electron is brought back to its initial state $|\downarrow\rangle$. Indeed, during the whole time evolution the system practically remains in the subspace spanned by $|\downarrow 10\rangle$ and $|\uparrow 10\rangle$ and all the other basis states are not visibly populated.}
	\label{fig:ZZZ}
\end{figure}

\subsection{ZZ Gate}
We start by a demonstration of the ZZ gate $\exp(i \frac{\pi}{4}  \sigma_z \otimes \sigma_z)$ which is part of the native gate set of our system and thus does not require decomposition. Instead, it is an entangling gate that can be achieved directly by optimizing the microwave pulse according to the method outlined in Section~\ref{section:gate_design}. We consider the gate for the two qubits $^{15}N$ and $^{13}C_1$. We set the gate time to 3\,\textmu s, and generate the control pulse by optimization of Eq.~\eqref{eq:FoM_entangling_gate}. We numerically calculate the time evolution and the gate fidelity of $\mathfrak{F}=0.9997$. The time dynamics of the system for the initial state $|\downarrow 10\rangle$ is shown in Figure \ref{fig:ZZZ}. 
We can see that the electron is excited without altering the nuclear-spin (logical basis state) population and then brought back to the initial state. However, the (partial) excitation leads to a geometric phase accumulated by each logical basis state. In this way the phase gate on the logical states is realized.

\subsection{Two-qubit Unitaries}
In this section, we test our approach by applying it to find sequences of control pulses that generate a CNOT gate and a SWAP gate between two nuclear spins, respectively. As we will see, the SWAP gate needs a longer circuit and we thus compare the results from the two different versions of the decomposition algorithm, sections~\ref{sec:decomposition_algorithm} and~\ref{sec:variant_two_qubit}. For all three cases we show the circuit diagrams and the time evolution of the system starting in one specific basis state. For all examples we choose the gate time of the entangling gates to be 3\,\textmu s, while we adjust the gate time of the nuclear spin gates to each problem: the gate time for each $\pi$ rotation is $T_N$; the gate time for each partial rotation by an angle $\zeta$ (i.e., $\exp (i \zeta \sigma_{x/y})$) is chosen as $\frac{\zeta}{\pi}T_N$ and the value of $T_N$ is gven below for each example.

\subsubsection{CNOT}
\begin{figure}[htb!]
	\centering
	\includegraphics[width = 0.45\textwidth]{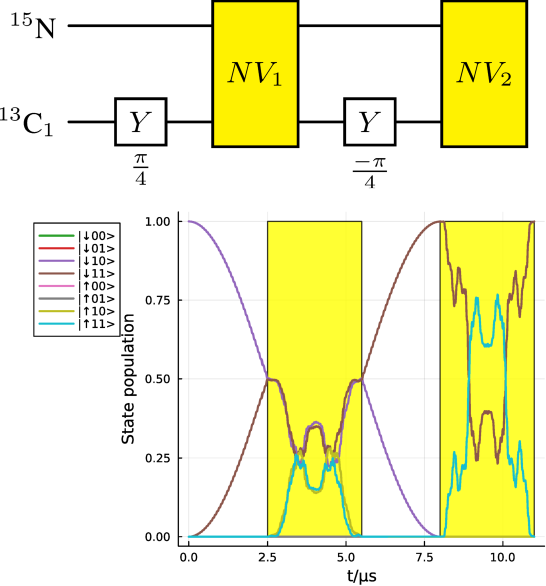}
	\caption{Decomposition of CNOT. The upper panel shows the circuit diagram of CNOT decomposed into the native gate set, where $NV_1 = \mathrm{diag}(1,1,e^{i\pi/2},e^{-i\pi/2})$ and $NV_2 = \mathrm{diag}(1,1,e^{-i\pi/2},e^{-i\pi/2})$. The lower panel shows the
		population evolution for the CNOT gate between the two nuclear spin qubits. The yellow boxes highlight the time intervals in which the entangling gates are applied. The system starts in a state with nuclear spins in state 
		$\vert 10\rangle$ and the electron in $\vert \downarrow\rangle$. After a CNOT, we observe the expected final nuclear spin state $\vert 11 \rangle$. Throughout the evolution of system, the electron is in a superposition only during the entangling gate times that take 3\,\textmu s each, reducing thus the loss of coherence via potential dephasing of the electron spin. For the CNOT gate we obtain a fidelity of $\mathfrak{F}=0.99992$.}
	\label{fig:CNOT}
\end{figure}

Via the decomposition algorithm of Section~\ref{sec:decomposition_algorithm}, the CNOT gate is decomposed into two diagonal gates and one pair of local rotations. The resulting circuit is shown in the upper panel of Figure \ref{fig:CNOT},~\cite{Kay2023}. The decomposition is
$$
CNOT = \begin{bmatrix}
	1 & 0 & 0 & 0\\
	0 & 1 & 0 & 0\\
	0 & 0 & 0 & 1\\
	0 & 0 & 1 & 0
\end{bmatrix} = K_0 (K_1 H K_1^{\dagger}),
$$ where $K_1 = \mathbbm{1}$ and
$$
K_0 = \exp\left(i \begin{bmatrix}
	0 & 0 & 0 & 0\\
	0 & 0 & 0 & 0\\
	0 & 0 & -\frac{\pi}{2} & 0\\
	0 & 0 & 0 & -\frac{\pi}{2}
\end{bmatrix}\right) \,,
$$
$$
H = \ket{0}\rangle\langle 0 \vert \otimes \mathbbm{1} + \vert 1\rangle\langle 1 \vert \otimes R_x\left(\pi\right).
$$
By noting that $R_x(\pi) = R_y(-\frac{\pi}{2}) R_z(\pi)R_y(\frac{\pi}{2})$ (see step 3. of the algorithm), we can rewrite
$$
H = (\mathbbm{1} \otimes R_y\left(-\frac{\pi}{2}\right)) \exp\left(i\begin{bmatrix}
	0 & 0 & 0 & 0\\
	0 & 0 & 0 & 0\\
	0 & 0 & \frac{\pi}{2} & 0\\
	0 & 0 & 0 & -\frac{\pi}{2}
\end{bmatrix}\right)(\mathbbm{1}\otimes R_y\left(\frac{\pi}{2}\right))\,.
$$
The protocol thus decomposes the CNOT between the two nuclear spin qubits into two entangling phase gates between the two qubits and two $\pi/2$-$y$ gates on the $^{13}C_1$ spin. Note that the entangling gates are driven via the microwave and designed to be fast compared to typical dephasing times, while during the slow single-qubit gates (driven by radio-frequency pulses) the electron spin is not in a superposition state.

The lower panel of Figure~\ref{fig:CNOT} shows the time evolution of the $|10\rangle$ basis state to the $|11\rangle$ state. The electron is initially in $|\downarrow\rangle$. We can indeed see that the electron is excited only during the two entangling gates, and after such gates, it always returns to the initial state. For the single-qubit gates we choose $T_N=10$\,\textmu s (so that each of the two single-qubit gates take $2.5$\,\textmu s). The total gate time for the CNOT is $11$\,\textmu s and the fidelity for the CNOT is $\mathfrak{F}=0.99992$.

\subsubsection{SWAP}
\begin{figure}[h]
	\centering
	\includegraphics[width = 0.45\textwidth]{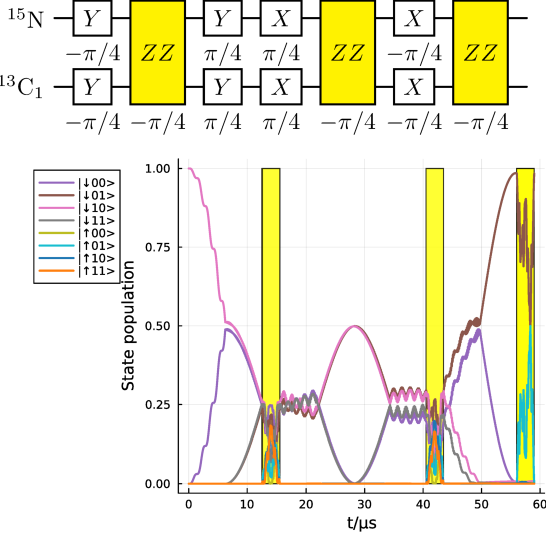}
	\caption{The upper panel shows the circuit diagram for the SWAP gate, where the $SU(4)$ unitaries are decomposed according to the variant of the algorithm from Section~\ref{sec:variant_two_qubit}.
		The lower panel shows the population evolution corresponding to the circuit. The yellow boxes highlight the time intervals in which the entangling gates are applied. The system starts in the state $\vert 10\rangle$ and evolves to  $\vert 01 \rangle.$ In the entire duration of the unitary, the electron is only excited during the short nuclear entangling gates (3\,\textmu s) and always returns to the initial state afterwards, ensuring robustness of the system against decoherence. Overall, for the SWAP gate we obtain a fidelity of $\mathfrak{F}=0.991$.}
	\label{fig:cano_SWAP}
\end{figure}

To decompose the SWAP gate into the native gate set of our approach, we have two choices, the main version of the algorithm, Section~\ref{sec:decomposition_algorithm} and the variant, Section~\ref{sec:variant_two_qubit}. We first use the variant for the SU(4) decomposition algorithm. The nuclear SWAP gate is thus decomposed to the sequence given in the upper panel of Figure~\ref{fig:cano_SWAP}, totaling 11 native gates, 3 entangling gates and 8 single-qubit gates.  The nuclear $-\frac{\pi}{4}$ "ZZ" gate is $\exp(i \frac{\pi}{4} \sigma_z \otimes \sigma_z)$.

We generate the pulses for all these 11 gates and show the time dynamics in the lower panel of Figure \ref{fig:cano_SWAP}. Again, the electron is excited only during the two entangling gates, and after such gates, it always returns to the initial state. For the single-qubit gates we choose $T_N=25$\,\textmu s. The total gate time for the SWAP is $59$\,\textmu s and the fidelity for the SWAP gate is $\mathfrak{F}=0.991$.

\begin{figure}[h]
	\centering
	\includegraphics[width = 0.45\textwidth]{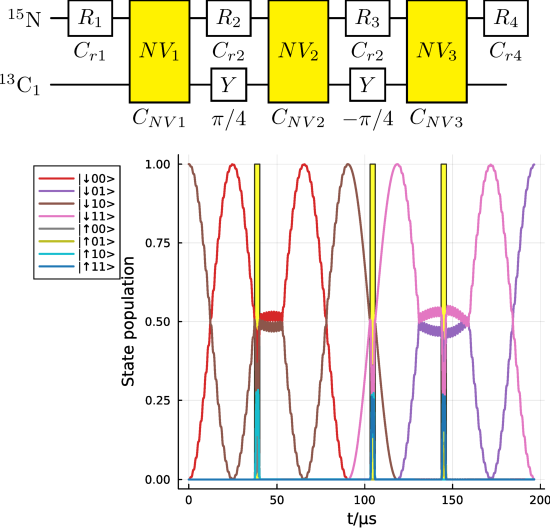}
	\caption{Decomposition of SWAP. The upper panel shows the circuit diagram for the nuclear SWAP gate according to the main version of the decomposition algorithm.  The lower panel shows the corresponding population evolution. The yellow boxes highlight the time intervals in which the entangling gates are applied. The system evolves from the initial state of $\vert 10 \rangle$ to $\vert 01\rangle$, according to the control pulses generating the SWAP gate. Overall, for the SWAP gate we obtain a fidelity of $\mathfrak{F}=0.990$.}
	\label{fig:SWAP}
\end{figure}

We now use the main version of the decomposition algorithm. The nuclear-nuclear SWAP gate is now decomposed to the sequence given in the upper panel of Figure~\ref{fig:SWAP}, this time totaling only 9 native gates, 3 entangling gates and 6 single-qubit gates. The gate parameters are given in appendix~\ref{appendix: C}, Table~\ref{tab:param_SWAP}.

We generate the pulses for all these 9 gates and show the time dynamics in the lower panel of Figure \ref{fig:SWAP}. Again, the electron is excited only during the three entangling gates, and after such gates, it always returns to the initial state. Since the circuit is slightly more complicated due to the different single-qubit gates, we have to choose a longer single-qubit gate time of $T_N=50\mu s$ to obtain a similar fidelity as for the decomposition according to Fig.~\ref{fig:cano_SWAP} (for $T_N=25\mu s$, here we obtain a fidelity of only $\mathfrak{F}=0.96$). The total gate time for the SWAP is now $196.5\mu s$ and the fidelity for the SWAP is $\mathfrak{F}=0.990$.

Comparing the two decomposition schemes for the nuclear SWAP gate into native gates, the variant of the algorithm involves standard gate angles such as $\pi/4$, whereas the approach using our main algorithm involves fewer but more general nuclear rotations. Since we have chosen to generate all single-qubit gates via resonant pulses with constant amplitude, we have to decompose the more general rotations into several rotations along $\sigma_x$ and $\sigma_y$. As a consequence, the variant of the algorithm leads to higher fidelities for the SWAP gate. In principle, however, one could choose to generate the single-qubit gates with QOC, in which case the decomposition with a smaller total number of gates could lead to an advantage.

\subsection{Three-qubit Unitaries: QFT}
\begin{figure*}[ht]
	\centering
	\includegraphics{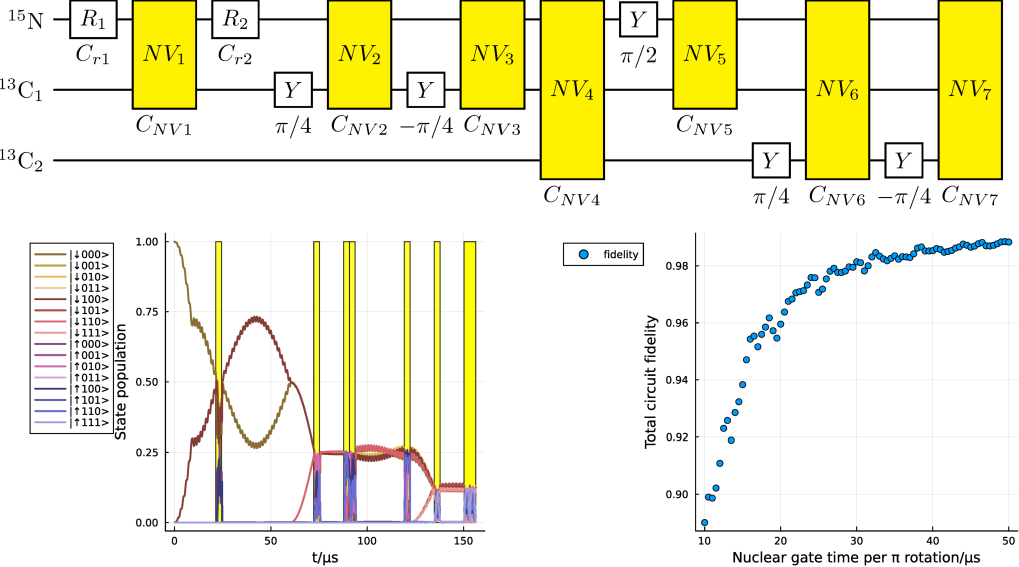}
	\caption{Decomposition of QFT3. The upper panel shows the circuit diagram of QFT3, single nuclear rotations about arbitrary axis are denoted by numbered $R_i$ with the rotation coefficients $C_{ri} = \zeta/2 \hat{n}$. The NV gates are labeled by $NV_i$ and the diagonal coefficients of the NV gates are given for each gate as $C_{NVi}$. The coefficients are listed in Table~\ref{tab:param_QFT3} in the Appendix.
	The lower left panel shows the corresponding time evolution for the initial state $\vert \downarrow 000 \rangle$. The yellow boxes highlight the time intervals in which the entangling gates are applied. As expected, it evolves into a superposition state, where all 8 three-nuclear states have a population of $0.125$ and the electron is set back to its initial state $|\downarrow\rangle$. The overall fidelity of the unitary evolution compared to the ideal circuit is $\mathfrak{F}=0.988$.
	The lower right panel shows the overall circuit fidelity for QFT3 versus the nuclear unit gate time $T_N$. With the pulses staying absolutely the same for the entangling gates, we choose the gate time for each rotation $\exp (i \zeta \sigma_{x/y})$ as $\frac{\zeta}{\pi}T_N$. The plot shows the fidelity of QFT3 over $T_N$.
	}
	\label{fig:QFT3}
\end{figure*}

For three-qubit QFT (QFT3), the decomposition algorithm gives a circuit consisting of 14 native gates, see upper panel of Figure~\ref{fig:QFT3} for the circuit diagram and Table~\ref{tab:param_QFT3}  of appendix~\ref{appendix: C} for the parameters of each gate in the circuit. There are three three-qubit gates, five two-qubit gates and seven single-qubit gates.

We again generate the pulses for all the 14 gates and simulate the time evolution (see lower left panel of Figure~\ref{fig:QFT3}). The fidelity of the circuit implementation is $\mathfrak{F}=0.988$.
The fidelity here is limited by the single-qubit fidelity as will be discussed in the following paragraph. To reach better fidelities, one needs substantially longer single-qubit gates or one needs to improve their fidelity by QOC.

The entangling gates are chosen to be optimized at a gate length of  3\,\textmu s (Note that NV7 is synthesized by two gates of half the rotation angle, leading to a total gate time of 6\,\textmu s). The nuclear gates, on the other hand, are first decomposed with the XYX-decomposition and than realized by constant pulses at resonance. In lower right panel of Figure~\ref{fig:QFT3}, we show the fidelity for QFT3 as a function of $T_N$. We can see that one needs at least $T_N=30$\,\textmu s to guarantee an overall fidelity of 90 $\%$, and for the lower left panel of Figure~\ref{fig:QFT3} we choose $T_N=50$\,\textmu s as a good compromise between circuit length and high fidelity. The total time to implement QFT3 is then $156.25$\,\textmu s.

In principle, the number of gates obtained from a circuit decomposition depends on the number of qubits and not on the circuit (see next section). Thus, a loosely connected circuit (that has a solution with considerably fewer gates) like QFT3 poses a challenge to the decomposition algorithm. However, with heuristics from Section~\ref{section:heur}, the overall decomposition sequence is reduced from 22 gates to 14 gates and thus provides a solid gateway to efficient QFT3 simulation. Due to the intrinsic complexity of the qubit connectivity graph for systems with more than two qubits, the decomposition algorithm cannot guarantee globally optimal decomposition solution. Additionally, an analysis of the time-optimality of the circuit also needs to take into account the pulse optimization and the length of the individual gates; shorter circuits still tend to be implementable in a shorter time and with higher fidelity.

\subsection{Beyond three qubits}
\begin{figure}[h]
\centering
\includegraphics[width = 0.45\textwidth]{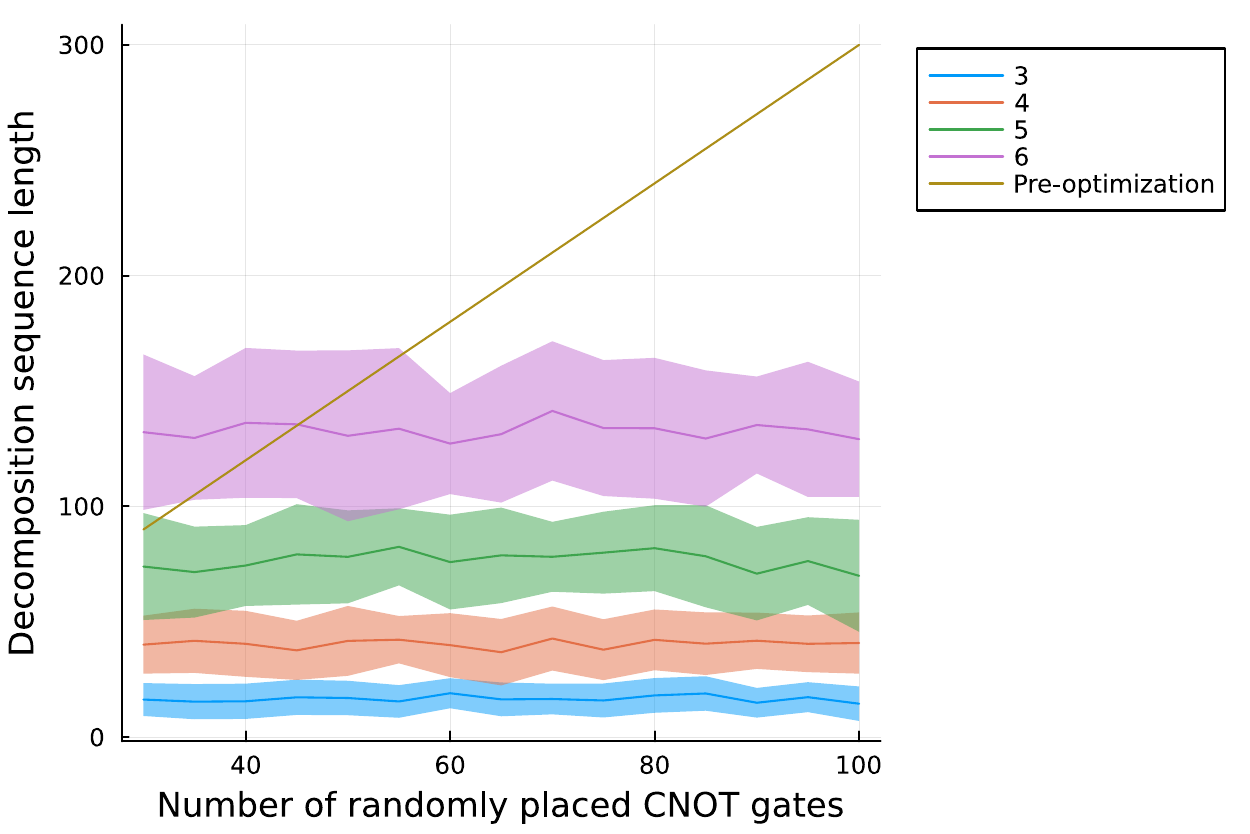}
\caption{Decomposition length against number of randomly placed CNOTs in the circuit. Each data set is for a specific number of qubits. As shown in the figure, while the naive synthesis method  (Pre-optimization) scales proportional to the number of CNOTs, the algebraic decomposition outputs an approximately constant circuit depth with respect to the number of CNOTs and only scales with the number of qubits present in the system, \textit{i.e.}, the size of the unitary.}
\label{fig:random_CNOTs}
\end{figure}

Here, we examine the ability of the decomposition algorithm to decompose random unitaries involving more than three qubits. We generate circuits with $N_{CNOT}$ two-qubit CNOT gates between randomly selected pairs of nuclear spins. Without decomposition, each such nuclear-nuclear CNOT gate could in principle be replaced by three CNOT gates between one of the nuclear spins and the electron or by the four gates found in Figure~\ref{fig:CNOT}. For each such instance, the naive circuit depth is thus approximately $ 3 N_{CNOT}$. Instead, the decomposition depth scales with the number of qubits in the circuit instead of the number of CNOTs. In Figure \ref{fig:random_CNOTs} we show how the decomposition circuit depth scales with the number of CNOTs randomly placed in the circuit and the number of qubits involved in the unitary. This is juxtaposed in the same graph with the naive synthesis method.
From this quantitative example, one can observe that the length of a decomposition sequence scales roughly with the number of qubits, \textit{i.e.}, the width of the circuit to be decomposed and stays roughly constant over the number of random CNOTs involved, \textit{i.e.}, the depth of the circuit.

\section{Discussion}\label{sec:discussion}
We have studied unitary decomposition and numerical simulation of the time dynamics of the involved gates, ranging from basic gates to single-, two- and three-qubit gates. Finally, the decomposition algorithm was tested for a random CNOT circuit for up to 6 qubits to analyze how the complexity of the decomposition result scales with circuit depth and circuit width (number of qubits involved). The sequence length after decomposition was found to be significantly smaller than before decomposition when a high level of connectivity between the qubits was reached. For circuits with sparse connections, the proposed method only yields sub-optimal solutions. However, due to the generality of the Cartan decomposition as a method, partial decomposition of the circuit or heuristic decomposition based on a decision tree can further optimize the decomposition results.
In Section~\ref{section:heur}, we have described a way to heuristically shorten the circuit and we have successfully tested it for QFT3. We have found that we can implement the resulting circuits with high fidelity within a realistic gate time. The gates we found via the decomposition algorithm often appear in the decomposition of not just one target unitary. For example, the $NV_1$ gate of the CNOT decomposition is the same as the $NV_{1-3}$ gates of the non-canonical SWAP gate decomposition. We have not considered any noise in the simulation and the dominant noise would be dephasing of the electron spin. Nonetheless, our gates involve superposition states of the electron spin only for a limited amount of time during the entangling gates. Furthermore, robustness against dephasing could be incorporated into the control pulses. Additionally, the single-qubit gates are generated with constant pulses and could be optimized by numerical methods for further speed-up if necessary, for instance, tools from machine learning to generate optimal control gates for a continuous gate set~\cite{Preti2022}.

\section{Conclusion and outlook}\label{sec:conclusions}
We have presented a systematic scheme for platform-native universal operations on a few-spin register for centrally-coupled qubit systems like NV centers in diamond. The proposed method first analyzes the native Lie algebra structure of the system Hamiltonian to find out the straightforwardly, or semi-analytically accessible gate set, and then use tailored Cartan unitary decomposition to translate any given target unitary into a sequence of native gates. The decomposition algorithm is light-weight and analytic, facilitating \textit{a priori} estimation of circuit complexity, while the pulse generation method is numerically converted to optimization on scalars, i.e., without matrix exponentiation. Furthermore, through analysis of the coherence time of the involved qubits, mediating entanglement of the nuclear spins through the central electronic spin (without using the electron as a qubit) supports higher coherence times for the entire circuit. 

As possible future research, other quantum computing platforms could be analyzed in a similar way for system-native universal computation. Conversely, it is also worth exploring the possibility to extend the set of native gates for the NV-center processor by adding different types of controls.

\begin{acknowledgments}
	The authors acknowledge fruitful discussion with M. Hendriks, J.Z. Bernad, F. Preti and R. Zeier.
	This work was supported by Germany’s Excellence Strategy – Cluster of Excellence Matter and Light for Quantum Computing (ML4Q) EXC 2004/1 – 390534769, and by the German Federal Ministry of Research (BMBF) under the project SPINNING (No. 13N16210), the European Union’s HORIZON Europe program via projects SPINUS (No. 101135699) and OpenSuperQPlus100 (No. 101113946) and by AIDAS-AI, Data Analytics and Scalable Simulation, which is a Joint Virtual Laboratory gathering the Forschungszentrum J\"ulich and the French	Alternative Energies and Atomic Energy Commission.
\end{acknowledgments}

\bibliography{ref}

\appendix
\section{Decomposition algorithms}
\label{appendix: A}
In Ref.~\cite{Nakajima2003}, the algorithm for decomposition of $G = K K_1 H K_1^\dagger$ is outlined, however, $K K_1$ and $K_1^\dagger$ are not necessarily elements of $SU(2^{N-1})$, so the recursion is not yet complete. Here, we provide the details of step 4 and 5 of Sec.~\ref{sec:decomposition_algorithm} for the further decomposition.
\begin{enumerate}
	\item  We want to decompose $K_l=K K_1$ and $K_r=K_1^\dagger$ using the involution
	$$\theta_1 (x) = \sigma_{N,x}\,x\,\sigma_{N,x}, $$ which produces the Cartan pair
	$$
	\mathfrak{m}_1 = \text{span}\left \{(i \id_{N-1} \oplus \mathfrak{su}(2^{N-1}))\otimes \sigma_z\right\}
	$$
	$$
	\mathfrak{k}_1 = \text{span}\left \{\mathfrak{su}(2^{N-1}) \otimes \id\right \}.
	$$
	with the corresponding maximal abelian Cartan subalgebra
	$$\mathfrak{h}_1 = i\text{span} \left \{\vert j \rangle \langle j \vert \otimes \sigma_z \right \},$$ which is intrinsically diagonal. We thus calculate $M^2 = \theta_1(K_{l,r}^\dagger) K_{l,r}$.
	\item We proceed to decompose $M^2$ by eigenvalue decomposition, such that $M^2 = QDQ^\dagger$, where $D$ is the diagonal matrix containing the eigenvalues of $M^2$ and $Q \in SU(2^N)$, i.e., $Q$ is a unitary similarity matrix ant its columns $\overrightarrow{q}$ are orthonormal eigenvectors of $M^2$. We aim to use the degree of freedom in the choice of these vectors such that
	\begin{itemize}
		\item the eigenvectors are mapped by $\overrightarrow{q} \mapsto \overrightarrow{q}'$ such that $(q_i')^{\dagger}q_j' = \Delta_{ij}$, i.e., the orthonormality and hence unitarity is conserved.
		\item the resulting unitary  $Q'$ satisfies the involution condition $\theta(Q') = \sigma_{N,z}Q'\sigma_{N,z} = Q'$, meaning that $Q' \in \exp(\mathfrak{k})$
		\item $Q'$ further satisfies the involution condition $\theta_1(Q') = Q'$, meaning that $Q' \in \exp(\mathfrak{k}_1)$, or equivalently, an element of $SU(2^{N-1})$
		\item with the conjugation of $Q'$, $H_1^2 = Q' M^2 Q'^\dagger$ is put into the correct form of an element in the Cartan subalgebra $\mathfrak{h}_1$
	\end{itemize} 
	\item We group the eigenvalues of $M^2$ and their corresponding eigenvectors into three categories. Since $M^2$ is unitary, the eigenvalues are either complex roots $e^{i\alpha}$ ($\alpha$ real) or $\pm 1$. Eigenvectors in each category span a subspace in the Hilbert space, so we can guarantee orthogonality globally provided orthogonality in each subspace. 
	\item Since $M^2$ is a special unitary, the complex eigenvalues come in complex conjugate pairs, \textit{i.e.} if $e^{i\alpha}$ is an eigenvalue of $M^2$ then $e^{-i\alpha}$ is also a valid eigenvalue. For eigenvalues $\pm 1$, even cardinality also applies. This means we can further group the eigenvalues and corresponding eigenvectors into pairs $M^2p_1=e_1 p_1$ and $M^2p_2=e_2 p_2$, where $e_1 = \large\frac{1}{e_2}$. We then form a new basis for the subspaces from these pairs.
	\begin{itemize}
		\item Given a pair $p_1$ and $p_2$, we form a new pair of basis vectors $q_1$ and $q_2$ by
		\item $q_1 = p_1 + \sigma_{N,z} p_1$
		\item $q_2 = \sigma_{N,x} q_1$
		and normalizing. (We treat the special cases of $q_1=0$ and $q_1=q_2$ in the subsection below.)
	\end{itemize}
	
	\item Now we can easily verify that the proposed mapping satisfies the following relations
	\begin{itemize}
		\item $\sigma_{N,z} q_1 = q_1$
		\item $\sigma_{N,z} q_2 = -q_2$
		\item $\sigma_{N,x} q_1 = q_2$
		\item $\sigma_{N,x} q_2  = q_1$
	\end{itemize}
	With this choice for the columns of $Q'$, the invariance under the involutions will indeed be satisfied: as $\sigma_{N,z} Q'$ essentially gives every even column a negative sign. One can check that this signs are eliminated again by the right multiplication of $\sigma_{N,z}$. For $\sigma_{N,x}$, every pair of odd and even columns are swapped twice, also leaving the $Q'$ invariant. Furthermore, by applying $M^2$ on the new eigenvectors we obtain
	\begin{equation}
		\begin{split}
			M^2 q_1 & = M^2 p_1 + M^2 (\sigma_{N,z} p_1)\\
			&=e_1 p_1 + \sigma_{N,z} (\sigma_{N,z} M^2 \sigma_{N,z}) p_1\\
			&=e_1 (p_1 + \sigma_{N,z} p_1)=e_1 q_1 \\
		\end{split}
	\end{equation}
	and
	\begin{equation}
		\begin{split}
			M^2 q_2& = M^2 \sigma_{N,x} p_1 + M^2 \sigma_{N,x} \sigma_{N,x} p_1\\
			&= \sigma_{N,x}(\sigma_{N,x} M^2 \sigma_{N,x}) p_1\\
			&\quad + \sigma_{N,x}(\sigma_{N,x} M^2 \sigma_{N,x}) \sigma_{N,z} p_1\\
			&=\sigma_{N,x} M^{2\dagger} p_1+\sigma_{N,x} M^{2\dagger} \left(\sigma_{N,z} p_1\right) \\
			&=e_1^*\left(\sigma_{N,x} p_1+\sigma_{N,x} \sigma_{N,z}p_1\right)=e_1^* q_2 \\
		\end{split}
	\end{equation}
	and hence
	$$
	{\left[\begin{array}{l}
			q_1 \\
			q_2
		\end{array}\right] M^2\left[\begin{array}{ll}
			q_1 & q_2
		\end{array}\right]=\left[\begin{array}{ll}
			e_1 & 0 \\
			0 & e_1^*
		\end{array}\right]=\exp \left(i \zeta \sigma_z\right)} \\
	$$
	resulting in $H_1^2$ with diagonal blocks. We have thus shown that under the new basis, $H_1^2 = Q' M^2 Q'^\dagger$ is indeed an element of $\mathfrak{h}_1$.
	\item In the final step we obtain $H_1$ by replacing $\zeta$ with $\frac{\zeta}{2}$ for all the pairs involved.
\end{enumerate}
\subsection{Orthogonality of manipulated eigenvectors}
So far we have implicitly assumed that all eigenvectors are orthogonal. Here, we examine this assumption more rigorously.
It is clear that eigenvectors corresponding to distinct eigenvalues are orthogonal to each other. However, when the unitary shows degeneracy in the spectrum, \textit{i.e.}, there exist multiple eigenvectors with the same eigenvalue, then cautious manipulation within the eigenspaces is needed to ensure the orthogonality of the transformed basis vectors that would later on form the unitary matrix that satisfies the involution conditions. Here, we analyze the possible structure of the eigenspaces and provide strategies to maintain the orthogonality of the transformed basis vectors, hence the unitarity of the matrix they form.

\begin{algorithm}[H]
	\label{alg}
	\begin{algorithmic}[1]
		\small
		\caption{Orthogonality filter.}
		\Function{Filter}{$\vec{v},\sigma, D, \vec{p},N$}
		\Comment{Where $\vec{v}$: list of non-singular eigenvectors, $\sigma$: involution operator, $D$: Dimension of matrix, $\overrightarrow{p}$: list of singular pairs, N: required number of basis vectors}
		\State Let $\epsilon$ be a small numerical near-zero threshold
		\For{$v$ in ${\vec{v}}$}
		\For{$p$ in $\overrightarrow{p}$}
		\If{$v^\dagger \sigma p > \epsilon$} 
		\Comment{If $v$ transformed not compatible with $p$ transformed}
		\State $\vec{v} = \vec{v}/v$
		\EndIf
		\EndFor
		\EndFor
		\State Let l be length of resulting $\vec{v}$
		\State Let V be D x l matrix s.t. V[:,i] = $\vec{v}_i$
		\State Let K = $V' \sigma V$
		\State Let $\overrightarrow{i} = [\,]$
		\For{$i = 1$ to $l$}
		\State n = $\Sigma_{j=1}^{l}(K[j,i] < \epsilon)$
		\If{$n >= N-1$}
		\State push!($\overrightarrow{i}$, $i$)
		\EndIf
		\EndFor
		\State Let K' = K[$\overrightarrow{i}$, $\overrightarrow{i}$]
		\State Let $l'$ = length($\overrightarrow{i}$)
		\For{$j = 1$ to $l'$}
		\For{$\overrightarrow{k}$ in Combinations(K'[j,j+1:end] $< \epsilon$) .+ 1}
		\State Let R = K'[$\overrightarrow{k}$,$\overrightarrow{k}$]
		\For{$t = 1$ to $N-1$}
		\State R[t,t] = 0
		\EndFor
		\If{$\Sigma_{a,b} R[a,b] < \epsilon$}
		\State Return $\vec{v}[\overrightarrow{i}[j;\overrightarrow{k}]]$
		\EndIf
		\EndFor
		\EndFor
		\EndFunction
	\end{algorithmic}
\end{algorithm}

First of all, under conjugation of the involution operator $\sigma$ (either $\sigma_{N,z}$ or $\sigma_{N,x}$) any eigenvector $v$ of $M$ with eigenvalue $e$ becomes a new eigenvector of the same matrix $M$ corresponding to an eigenvalue $e^*$:
	$$M \sigma v = \sigma \sigma M \sigma v = \sigma M^{\dagger} v = \sigma e^* v.$$

As a consequence, eigenvectors from distinct eigenspaces remain orthogonal to each other under the transformation considered above. 
\begin{lemma}
	Let $v_i$ and $v_j$ be two eigenvectors with distinct eigenvectors, such that $v_i^\dagger v_j = 0$. Then according to the transformation mentioned above, new basis vectors are constructed from them, denoted by $u_i$ and $u_j$:
	$$u_i = v_i \pm \sigma v_i, u_j = v_j \pm \sigma v_j,$$
	such that $u_i^\dagger u_j = 0.$
\end{lemma}
The proof of the lemma is as follows:
$$u_i^\dagger u_j = (v_i \pm \sigma v_i)^{\dagger} (v_j \pm \sigma v_j) = 2v_i^\dagger v_j + 2 v_i^\dagger \sigma v_j = 2 v_i^\dagger \sigma v_j.$$ 
Since $\sigma v_j$ is itself an eigenvector with eigenvalue that is the complex conjugate of that of $v_j$, as long as this eigenvalue is distinct from that of $v_i$, the last expression is $0$ and the two vectors $u_i$ and $u_j$ are orthogonal.

For both the involutions used in the algorithm, the first basis vector is taken to be $u = v \pm \sigma v.$ However, the problem of singular normalized $u$ occurs when $v = \pm \sigma v$. This can happen for eigenvectors from all three eigenspaces, $\pm 1$ and complex. The solution to the singularity problem is the following procedure.

For each eigenspace separately, if there exists pairs of singular eigenvectors, within one pair $v_1, v_2$, $v_1 = \sigma v_1, v_2 = -\sigma v_2,$ then for the first involution under $\sigma_{N,z}$ let the transformed $u_1, u_2$ simply be $u_1 = v_1, u_2 = v_2.$ And under the second involution $\sigma_{N,x},$ $u_1 = v_1, u_2 = \sigma_{N,x} v_1$. One can easily check that the new transformed basis vectors are orthogonal and satisfies the respective involution conditions.

Once the pairs of singular vectors are incorporated into the new pool of basis vectors, the remaining non-singular basis vectors need to be filtered by an orthogonality algorithm to ensure unitarity of the resulting matrix.

For eigenspaces with eigenvalues $\pm 1$, the selection procedure is outlined in Algorithm \ref{alg}. For eigenspaces of complex eigenvalue pairs, for each pair of eigenvalue $e$ and $e^*$, one can simply choose all the vectors from either $e$ or $e^*$ since the involution-conjugated vectors are by default eigenvectors with eigenvalue complex conjugated, thus orthogonal to the original eigenvectors. Usually, each complex subspace is small enough that non-orthogonality is highly unlikely. However, one can enforce this condition by putting the selected set of eigenvectors through the orthogonality filter of Algorithm \ref{alg}.

\section{Parameter table}
\label{appendix: C}
Here, we provide the parameter table with the numerical values of each gate in the decomposition of the non-canonical SWAP circuit (Table~\ref{tab:param_SWAP}) as well as in the decomposition of the QFT3 circuit (Table~\ref{tab:param_QFT3}).

\begin{table}[H]
	\caption{Parameter table including numerical values for gates in non-canonical SWAP decomposition sequence, corresponding to the circuit diagram in Figure \ref{fig:SWAP}, following the same convention of notation as in the previous case. Note: The parameter A appearing in the  table has the value $\frac{2\pi}{3\sqrt{3}}$ up to numerical precision.}
	\label{tab:param_SWAP}
	\centering
	\begin{tabular}{|c|c|}
		\hline
		Label  & Coefficients\\
		\hline
		C$_{r1}$ & 0, $-3\pi/4$, 0.0\\
		\hline
		C$_{NV1}$ & 0, 0, $\pi/2$, $-\pi/2$\\
		\hline
		C$_{r2}$ & -A, A, -A\\
		\hline
		C$_{NV2}$ & 0, 0, $\pi/2$, $-\pi/2$\\
		\hline
		C$_{r3}$ & A, -A, -A\\
		\hline
		C$_{NV3}$ & 0, 0, $\pi/2$, $-\pi/2$\\
		\hline
		C$_{r4}$ & -A, A, -A\\
		\hline
	\end{tabular}
\end{table}

\begin{table}[H]
	\caption{Parameter table including numerical values for gates in QFT3 decomposition sequence. For nuclear rotations, the three coefficients labelled by $C_{ri}$ correspond to a rotation gate $\exp(i (C_{ri, 1}\sigma_x + C_{ri,2} \sigma_y + C_{ri,3} \sigma_z))$ and the $2^m$ coefficients listed for NV diagonal gates are directly the diagonal coefficients in $\exp(i \text{diag}(C_{NVi}))$, where m is the number of qubits involved.}
	\label{tab:param_QFT3}
	\centering
	\begin{tabular}{|c|c|}
		\hline
		Label  & Coefficients\\
		\hline
		C$_{r1}$ & -0.4576,0.6849, -0.4576\\
		\hline
		C$_{NV1}$ & $\pi/4$, -$\pi/4$, 0, 0\\
		\hline
		C$_{r2}$ & $\pi/(2\sqrt{2})$,$\pi/(2\sqrt{2})$, 0.0\\
		\hline
		C$_{NV2}$ & $\pi/4$, -$\pi/4$, $\pi/4$, -$\pi/4$ \\
		
		\hline
		C$_{NV3}$ & $-\pi/2$, $\pi/4$, $-\pi/2$, $3\pi/4$\\
		\hline
		C$_{NV4}$ & $\pi/8$,$-\pi/8$,$-\pi/8$, $\pi/8$, $\pi/4$,-$\pi/4$,0,0\\
		
		\hline
		C$_{NV5}$ & 0,0,$-\pi/2$, $\pi/2$\\
		
		\hline
		C$_{NV6}$ & $\pi/4$, $-\pi/4$, $\pi/4$, $-\pi/4$, $\pi/4$, $-\pi/4$, $\pi/4$, $-\pi/4$\\
		
		\hline
		C$_{NV7}$ & $-\pi/2$, $-\pi$, $-\pi/2$, $-\pi$, $\pi/2$, 0, $\pi/2$, 0\\

		\hline
	\end{tabular}
\end{table}

\section{Hamiltonian for Group IV defects}
\label{app:groupiv}
For systems where the electron has spin 1/2 we have the magnetic states $m_s=-1/2$  ($|\downarrow\rangle$) and $m_s=1/2$ ($|\uparrow\rangle$). The static Hamiltonian is then
\begin{eqnarray}
	H_{\mathrm{static}}=\frac{\omega_{el}}{2}\sigma_z + 
	\sum_i  \left(- \gamma_i B_0  + \frac{A_{i,zz}}2 \sigma_z \right)I_{i,z}\,.
\end{eqnarray}
The single-qubit Hamiltonian 
\begin{eqnarray}
	\hat{H}_{SQ}(t)=-\exp(i H_{\mathrm{static}}t)\sum_i \gamma_i B_1(t) I_{i,x} \exp(-i H_{\mathrm{static}}t)\\
	= -|\downarrow\rangle\langle \downarrow |\sum_i (I_{i,x} \cos (\omega_i^+ t) + I_{i,y} \sin (\omega_i^+t) ) \gamma_i B_1(t)\nonumber\\
	-|\uparrow\rangle\langle \uparrow |\sum_i (I_{i,x} \cos (\omega_i^- t) + I_{i,y} \sin (\omega_i^- t) ) \gamma_i B_1(t)\\
	\approx -|\downarrow\rangle\langle \downarrow |\sum_i (I_{i,x} \cos (\omega_i^+ t) + I_{i,y} \sin (\omega_i^+t) ) \gamma_i B_1(t)
\end{eqnarray}
with $\omega_i^+ =\gamma_i B_0  + \frac{A_{i,zz}}2$ and $\omega_i^- = \gamma_i B_0  - \frac{A_{i,zz}}2$ takes a form very similar to Eq.~\eqref{eq:single_qubit_ham}.

For the multi-qubit Hamiltonian, we obtain

\begin{eqnarray}
	\hat{H}_{NV} &=& \exp(i H_{\mathrm{static}}t)\frac{\gamma_{NV}}{\sqrt 2} B_1(t) \sigma_x\exp(-i H_{\mathrm{static}}t)\\
	&=&\large\frac{\gamma_{NV}}{\sqrt 2} B_1(t) \sum_{l=1}^{2^N}  \big[
	\cos \left(\omega_{el} t+\sum_{j=1}^N F(s_{lj})\beta_j t\right)\sigma _ { x }\nonumber\\
	&&- \sin \left(\omega_{el} t+\sum_{j=1}^N F(s_{lj}) \beta_j t\right) \sigma_y\big]\otimes \ket{l}\!\bra{l}  \nonumber\\
	&=&\large\frac{\gamma_{NV}}{\sqrt 2} B_1(t) \sum_{l=1}^{2^N}  \big[
	\cos \left(\omega_{el} t+\tilde\omega_l t\right)\sigma _ { x }\nonumber \\
	&&- \sin \left(\omega_{el} t+\tilde\omega_l t\right) \sigma_y\big]\otimes \ket{l}\!\bra{l}  \,,
\end{eqnarray}
which is the same form as Eq.~\eqref{eq:multi_qubit_ham}, but with $F(0)=-1/2$ and $F(1)=1/2$ and thus slightly different $\tilde{\omega}_l=\sum_{j=1}^N F(s_{lj})\beta_j$.

\end{document}